# Odd-parity linear magnetoresistance and the planar Hall effect

Yishu Wang[1,2], Patrick A. Lee[1,3], D. M. Silevitch[1], F. Gomez[1], S. E. Cooper[4], Y. Ren[5], J.-Q. Yan[6], D. Mandrus[6,7], T. F. Rosenbaum[1,*], Yejun Feng[1,4,*]

[1]Division of Physics, Mathematics, and Astronomy, California Institute of Technology, Pasadena, California 91125, USA

[2]The Institute for Quantum Matter and Department of Physics and Astronomy, The Johns Hopkins University, Baltimore, Maryland 21218, USA

[3]Department of Physics, Massachusetts Institute of Technology, Cambridge, Massachusetts 02138, USA

[4]Okinawa Institute of Science and Technology Graduate University, Onna, Okinawa 904-0495, Japan

[5]The Advanced Photon Source, Argonne National Laboratory, Argonne, Illinois, 60439, USA

[6]Materials Science and Technology Division, Oak Ridge National Laboratory, Oak Ridge, Tennessee 37831, USA

[7]Department of Materials Science and Engineering, University of Tennessee, Knoxville, Tennessee 37996, USA

*Corresponding authors. Emails: tfr@caltech.edu, yejun@oist.jp

**One sentence summary:**

The traditional parity-delineated boundary between magnetoresistance and the Hall effect is overturned when time reversal symmetry is naturally broken.

**Abstract:**

**The phenomena of odd-parity magnetoresistance and the planar Hall effect are deeply entwined with ferromagnetism. The intrinsic magnetization of the ordered state permits these unusual and rarely observed manifestations of Onsager's theorem when time reversal symmetry is broken at zero applied field. Here we study two classes of ferromagnetic materials, rare-earth magnets with high intrinsic coercivity and antiferromagnetic pyrochlores with strongly-pinned ferromagnetic layers at domain walls, which both exhibit odd-parity magnetoresistive behavior. The peculiar angular variation of the response with**

**respect to the relative alignments of the magnetization, magnetic field, and current reveal the two underlying microscopic mechanisms: spin-polarization-dependent scattering of a Zeeman-shifted Fermi surface and magnetoresistance driven by the anomalous velocity physics usually associated with the anomalous Hall effect.**

**TEXT:**

The study of galvanomagnetic effects suffuses condensed matter, material, and device physics, tracing its origins to the pioneering work of Lord Kelvin and Edwin Hall in the 19[th] century. The response of electrons to applied magnetic and electric fields has revealed the fundamental characteristics of the Fermi surface in metals, stimulated paradigms of electron correlations from Mott to Kondo, and enabled devices from spintronics [1-2] to topological circuitry [3]. However, the link between macroscopic response and microscopic origins can be obtuse, relying upon the nuanced parsing of functional forms (negative vs positive, linear vs. quadratic, angular modulation, etc.) over a controlled parameter space [4-6].

One of the first rules of magnetotransport for experimentalists is to separate the diagonal and off-diagonal components of the resistivity tensor by testing for even vs. odd parity under field reversal. The magnetoresistance (MR), $\rho_{xx}(H)$, is even; the Hall resistance, $\rho_{xy}(H_z)$, is odd. Here we explore the phenomenon of odd-parity MR, rarely established experimentally [7-11] because of the usual symmetry assumptions. In a ferromagnet, an intrinsic magnetization **M** breaks time reversal symmetry even at field **H**=0, and the Onsager's relation for electrical conductivity $\sigma$ takes the form $\sigma_{ij}(\mathbf{H},\mathbf{M}) = \sigma_{ji}(-\mathbf{H},-\mathbf{M})$ [8, 11]. The salient requirement separating odd and even-parity MR is whether **M** remains constant through a reversal of the measurement field. With fixed **M**, we identify two kinds of odd parity contributions to the electrical current:

$$\mathbf{j}=A(\mathbf{M}\cdot\mathbf{H})\mathbf{E} + B(\mathbf{M}\times\mathbf{E})\times\mathbf{H} \qquad (1)$$

where *A* and *B* are constants, and **E** is the electrical field. We refer to the first term as odd parity MR. The second term, $(\mathbf{M}\times\mathbf{E})\times\mathbf{H}$, is equivalent to $(\mathbf{M}\cdot\mathbf{H})\mathbf{E} - (\mathbf{E}\cdot\mathbf{H})\mathbf{M}$, so that it includes both the odd-parity MR and a new term, $(\mathbf{E}\cdot\mathbf{H})\mathbf{M}$, which we shall refer to as the odd-parity planar Hall effect. As we demonstrate below, removing the requirement that **M** be parallel to **H** opens the door for a wide range of odd-parity galvanomagnetic behavior, encompassing both transverse and longitudinal MR [12] and the magnetic planar Hall effect [13].

We recall that Onsager's relation for non-magnetic materials gives $\sigma_{ij}(\mathbf{H}) = \sigma_{ji}(-\mathbf{H})$. In the case of an isotropic system, the electron transport equation [14] can be expanded to linear and quadratic orders of **E** and **H**, respectively, as $\boldsymbol{j} = (\sigma_{xx}^{(0)} + \beta H^2)\mathbf{E} + \sigma_{xy}^{(0)}\mathbf{E}\times\mathbf{H} + \gamma(\mathbf{E}\cdot\mathbf{H})\mathbf{H}$, where $\sigma_{xx}^{(0)}$ is the zero-field conductivity, $\sigma_{xy}^{(0)}$ is the Hall conductivity, and $\beta$ and $\gamma$ are numerical coefficients. The last term, $(\mathbf{E}\cdot\mathbf{H})\mathbf{H}$, leads to even-parity quadratic galvanomagnetic behavior in both the anisotropic MR and the planar Hall effect [15].

We proceed with a concrete phenomenological description to demonstrate the microscopic origins of Eq. 1, similarly starting from a transport equation for charge carriers of velocity $\boldsymbol{v}$ in the presence of a constant **M**, independent of **H**:

$$\left(\frac{d}{dt} + \frac{1}{\tau}\right)\boldsymbol{v} = \frac{e}{m}\mathbf{E} + \frac{e}{mc}\boldsymbol{v}\times\mathbf{H} + \alpha\mathbf{M}\times\mathbf{E}, \qquad (2)$$

with $\tau$ the typical relaxation time between scattering events, $e$ the electron charge, $m$ the mass of the electron, $c$ the speed of light, and $\alpha$ a numerical coefficient. The first two terms on the right-hand side represent the transport process driven by the externally applied field **E** and the Lorentz force, respectively, while the last term captures the anomalous Hall effect due to the presence of ferromagnetism. Solving $\boldsymbol{v}$ iteratively to first order in $H$ for a steady state solution gives:

$$\boldsymbol{j} = ne\boldsymbol{v} = \sigma_{xx}^{(0)}\mathbf{E} + \sigma_{xy}^{(0)}\mathbf{E}\times\hat{\mathbf{H}} + \sigma_{xy}^{(A)}\mathbf{E}\times\hat{\mathbf{M}} + \sigma_{xx}^{(0)}\frac{e\alpha}{c}(\mathbf{M}\times\mathbf{E})\times\mathbf{H}, \qquad (3)$$

where $\sigma_{xx}^{(0)} = ne^2\tau/m$, $\sigma_{xy}^{(0)} = \frac{ne^3\tau^3}{m^2 c}H$, and $\sigma_{xy}^{(A)} = -en\alpha\tau M$.

The final term, $(\mathbf{M}\times\mathbf{E})\times\mathbf{H} = (\mathbf{M}\cdot\mathbf{H})\mathbf{E} - (\mathbf{E}\cdot\mathbf{H})\mathbf{M}$, yields a general mechanism for obtaining both odd-parity linear magnetoresistance and the planar Hall effect, attributable to the same origin as the ferromagnetism-induced anomalous Hall effect (the third term in Eq. 3) [16]. Although odd-parity linear MR from this origin was theoretically discussed [11], as we will demonstrate below, a clear experimental verification relies on the planar Hall configuration. Various aspects of the planar Hall effect have been explored in the literature, but it is generally assumed that $H$ is large compared to the coercive field, so that **M** is parallel to **H.** The resulting term, $(\mathbf{E}\cdot\mathbf{M})\mathbf{M}$, is even under **H** reversal [13]. By contrast, we keep **M** and **H** independent so that the planar Hall term, $(\mathbf{E}\cdot\mathbf{H})\mathbf{M}$, is odd in **H**.

We note that the assumption of a single $\tau$ and $n$ in Eq. 3 breaks down in a ferromagnet [17], which leads to an additional mechanism for inducing odd-parity linear MR [8, 10]. Here, electrons of spin parallel (up) and antiparallel (down) to **M** have different scattering times $\tau_{up}$ and $\tau_{dn}$, with the carrier densities $n_{up}$ and $n_{dn}$ tuned linearly by the applied field through a Zeeman energy shift at the Fermi surface [17]. To first order in $H$, this brings a correction to $n\tau$ in $\sigma_{xx}^{(0)}$ where $n\tau$ is replaced by $n_{up}\tau_{up} + n_{dn}\tau_{dn} = \left((n_{up} - n_{dn})(\tau_{up} - \tau_{dn}) + (n_{up} + n_{dn})(\tau_{up} + \tau_{dn})\right)/2$, with $(n_{up} - n_{dn}) \sim H$. After projection along **M**, the first term in Eq. 3 gives a field-dependent MR which is negatively sloped and proportional to (**M·H**)**E**. With this correction term, Eq. 3 leads to the general form of Eq. 1. This linear form was observed in systems of saturated moments at high field [12]. Since the carrier density difference is linear in $H$, this argument would introduce a second-order in $H$ correction to the last term (**M×E**)×**H** in Eq. 3 and hence is neglected here.

We investigate experimentally the two odd-parity linear forms, (**M×E**)×**H** and (**M·H**)**E**, using physical systems with large coercivities of differing microscopic origins. In the rare-earth magnet SmCo$_5$, the combination of crystalline anisotropy along the *c*-axis and a needle-like microstructure [18] gives rise to coercivities up to ±2T at $T = 300$ K (Fig. S1). Additionally, we leverage the presence of parasitic ferromagnetism at metallic domain walls in insulating antiferromagnets [19-20], where the large coercivity stems from the ferromagnetic moments being pinned by the antiferromagnetic bulk below the Néel temperature, $T_N$. These conditions are met by pyrochlore-structured all-in-all-out antiferromagnets [9-10, 21-22] such as Eu$_2$Ir$_2$O$_7$ and Cd$_2$Os$_2$O$_7$ (Fig. S2), which can preserve constant ferromagnetic domain walls to at least ±9 T.

We first examine the galvanomagnetic behavior of several SmCo$_5$ samples in three configurations, transverse MR (Fig. 1A), longitudinal MR (Fig. 1B) and planar Hall (Fig. 1C), with both positive and negative magnetization. At $T = 300$ K, all of the $\Delta\rho(H)=\rho(H)-\rho(0)$ curves manifest an odd-parity linear behavior over a field range of ±5000 Oe with no noticeable quadratic component. The field range is chosen so that **M** remains essentially unchanged (Fig. S1). The oppositely-sloped resistivity with regard to either **M** or -**M** indicates that the initial magnetization is the determining factor for odd-parity, rather than the trivial experimental artifact of a projected ordinary Hall component.

To fully verify the functional forms (**M×E**)×**H** and (**M·H**)**E**, and to eliminate the possibility of cross-contamination between geometries, we made measurements over extended rotational degrees of freedom between all vectors **M**, **E** (as current **I**), and **H**, physically realized by a rotator stage inside our magnet/cryostat. We plot in Fig. 2 six sets of measurements, illustrating the resistive behavior of SmCo$_5$ in the various **M**–**E**–**H** configurations. All of the raw $\rho(H)$ curves are linear without discernable quadratic components, with the slopes d$\rho$(H)/dH in each configuration either null or sinusoidal in the rotation angle $\theta$. The first four sets of angular dependence, sinusoidal in Figs. 2A and 2B and null in Figs. 2C and 2D, are consistent with both (**M·H**)**E** and (**M×E**)×**H**. We separate the two terms in Figs. 2E and 2F. The sinusoidal form in Fig. 2E supports the presence of (**M·H**)**E**, given that (**M×E**)×**H** vanishes. Conversely, the magnetic planar Hall effect in Fig. 2F points to the presence of the (**M×E**)×**H** term, where (**M·H**)**E** is zero.

We note that contamination from other configurations is unlikely to explain the evidence for the planar Hall effect that emerges in Fig. 2F. A misalignment of the leads in the MR of Fig. 2B could in principle provide a nonzero projection of the planar Hall channel, but such a misalignment would phase shift the observed cos($\theta$) dependence towards a sin($\theta$) form. We also rule out mixing of effects from Fig. 2E, as the signal in Fig. 2F surpasses that of Fig. 2E by two orders of magnitude when normalized by the residual $\rho_{xz}$(H=0).

Compared to the bulk ferromagnetism of SmCo$_5$, ferromagnetic domain walls in antiferromagnetic Cd$_2$Os$_2$O$_7$, sketched in Fig. 3A, demonstrate linear odd-parity MR over a much larger field range, making such materials potentially more interesting for device applications. The measurements, however, are more technically challenging. With the magnetic field perpendicular to the sample surface, an imperfect four-lead geometry leads to a mixture of diagonal and off-diagonal terms, and as the domain walls are substantially more conductive than the bulk, the current paths are distorted, exacerbating this effect. The spatial inhomogeneity provides a realization of the Parish-Littlewood mobility fluctuation model [23], difficult to solve over a three-dimensional random domain structure [24]. Here, we use a van der Pauw (vdP) measurement geometry to highlight the mixture of the odd-parity linear behaviors, and separate the ordinary Hall and magnetic MR components at the macroscopic level by using the constant domain wall magnetization **M** as an independent variable.

Using a double rotation stage (Figs. 3B, 3C), we induced a constant ferromagnetic moment by cooling with the applied field parallel to the four-lead sample surface. After cooling below $T_N$ = 227 K, the magnetizing field was removed and the sample rotated by 90° so **H** becomes normal to the surface (Fig. 3C). The magnetization was prepared in a series of in-plane orientations $\phi$ by performing multiple cooldowns from $T$ = 300 K with differing orientations of the sample to the magnetizing field (Fig. S2). We find that the two independent MR channels and the two reciprocal Hall channels are all odd-parity and linear at $T$ = 195 K < $T_N$ (Fig. 3D), and their slopes oscillate with a 360° period in ϕ. The average linear slope of the Hall channels is ϕ-independent and equal to the ϕ-average slopes of the individual channels (Fig. 3F). As a check, we show in Fig. 3E that the linear slopes of the Hall channels are ϕ-independent and identical above $T_N$.

The ϕ dependence of the linear slope below $T_N$ (Figs. 3F, 3G) is explainable by the current-voltage reciprocal relationship, albeit under two different conditions. For ferromagnets, the reciprocal relationship states $R_{43,12}(\mathbf{H}, \mathbf{M}) = R_{12,43}(-\mathbf{H}, -\mathbf{M})$ (Fig. 3D inset, Ref. [25]). For a constant $\mathbf{M}$, the linear slopes of the two Hall channels at one ϕ position are not reciprocal. Instead, this relationship connects slopes measured at ϕ and ϕ + 180°, *i.e.* where $\mathbf{M}$ is reversed between two field-in-plane cooldowns. Additionally, at each ϕ position the opposite-phase components of the linear slopes of two reciprocal Hall configurations mirror the oppositely-valued ρ(*H*=0). Taken together, this explains the 360° period in ϕ instead of 180°, and indicates that the oscillating part of the linear slope is of magnetoresistive origin. As with the rare-earth ferromagnets, our results are consistent with origins in both (**M**·**H**)·**E** and (**M**×**E**)×**H**.

The normal voltage-current reciprocal relationship of $R_{43,12}(\mathbf{H}) = R_{21,43}(\mathbf{H})$ should hold for all ϕ for parts of the sample which either are nonmagnetic or have soft moments that follow **H**. This includes contributions from both normal electronic Hall and anomalous Hall from antiferromagnetic and/or paramagnetic spins [26-27], where canted spin moments vary linearly with applied field. Using the ϕ-dependence of linear slopes from a series of in-plane **M**, we are able to separate macroscopically the Hall and magnetoresistive components, despite both showing odd-parity linear behavior.

Across the Hall and two vdP MR configurations, ratios of the ϕ-averaged value and oscillation amplitude of the linear slope are not constant (Figs. 3F, 3G), suggesting differing origins for the two behaviors, as either a ϕ-independent Hall effect or a ϕ-oscillating MR. Further attribution of the odd-parity linear MR in $Cd_2Os_2O_7$ to either (**M**·**H**)·**E** or (**M**×**E**)×**H** origin is difficult, as both **M** and **E** contain components perpendicular and parallel to **H** along random domain walls. Additional clarity may be obtained by measuring thin film samples, reducing the difficulty of modeling domain wall conductance to two-dimensions [9, 23-24].

Any antiferromagnet with pinned ferromagnetic domain walls could exhibit transport characteristics that convolute contributions from the bulk and the domain wall surface, muddying interpretations of often coincident magnetic and metal-insulator phase transitions, such as those observed in $Cd_2Os_2O_7$ and other iridate pyrochlores. The identification of odd-parity MR arising solely from the ferromagnetic domain walls provides the means to characterize properly the intrinsic character of the bulk antiferromagnet and its evolution with temperature and field. It also informs design parameters for magnetic heterostructures with pronounced spin-orbit effects. A host of magnets with a large coercivity now become candidates for leveraging odd-parity MR and the planar Hall effect to elucidate their microscopic physics.

**Acknowledgments**
**Funding:** Y.F. acknowledges support from the Okinawa Institute of Science and Technology Graduate University (OIST), with subsidy funding from the Cabinet Office, Government of Japan. We also acknowledge the Mechanical Engineering and Microfabrication Support Section of OIST for the usage of shared equipment. The work at Caltech was supported by National Science Foundation Grant No. DMR-1606858. P.A.L. acknowledges support from the US Department of Energy, Basic Energy Sciences, Grant No. DE-FG02-03ER46076. J.-Q. Y. and D.M. acknowledge support from the US Department of Energy, Office of Science, Basic Energy Sciences, Division of Materials Sciences and Engineering. The work at the Advanced Photon Source of Argonne National Laboratory was supported by the US Department of Energy Basic Energy Sciences under Contract No. NEAC02-06CH11357. **Author contributions:** Y.W., Y.F., and T.F.R. conceived of the research; J.-Q.Y. and D.M provided osmate samples; Y.W., Y.F., D.M.S., F.G., S.E.C., and Y.R. performed experiments; P.A.L., Y.F., and Y.W. developed the theoretical interpretation; Y.W., Y.F., P.A.L., and T.F.R. prepared the manuscript. **Data and materials availability:** All data is available in the main text or the supplementary materials.

**List of Supplementary materials:**
Materials and Methods
Figs. S1 to S3

**Figure captions:**

**Fig. 1. Ferromagnetism induced odd-parity linear magnetoresistance and the planar Hall effect.** Using a fixed geometry relative to the field (schematics), $\Delta\rho(H)=\rho(H)-\rho(0)$ is plotted for three types of galvanomagnetic behavior, (A) transverse MR, (B) longitudinal MR, and (C) the planar Hall effect, relative to two different magnetic states **M** and **−M** in SmCo$_5$ samples. The magnetization is set by a field (±14 T) that is much larger than the measuring field (±0.5 T). In all three cases, odd-parity linear MR of opposite slopes were observed. The two-way switching behavior rules out trivial explanations arising from imperfections in the contact geometry, and pinpoints the origin in a magnetization that is independent of the measurement field.

**Fig. 2. Two mechanisms of odd-parity magnetoresistance and the planar Hall effect.** With two identified mechanisms to generate the odd-parity linear MR behavior, six rotational scenarios (A-F) were designed to create various geometries between **M**, **E**, and **H** in order to test the functional forms of both (**M·H**)**E** and (**M×E**)×**H**. In each figure, we plot schematics of the rotation scenario, raw data $\rho(H)$ at major angular positions, and either the normalized linear slope

$1/\rho_{xx}(H=0)$ d$\rho_{xx}$/d$H$ or the linear slope d$\rho_{xz}$/d$H$ vs. the rotation angle $\theta$. While panels (A)-(D) could arise from either (**M·H**)**E** or (**M**×**E**)×**H** terms, the angular dependence of the MR in (E) and the planar Hall effect in (F) are evidence of the former and latter functional forms, respectively. Similar odd-parity MR and rotation $\theta$-dependence was also observed in another highly coercive magnet Nd$_2$Fe$_{14}$B (Fig. S3).

**Fig. 3. Odd-parity linear magnetoresistance from ferromagnetic domain walls.** (A) A schematic view of the domain boundary along the <1,1,1> plane of Cd$_2$Os$_2$O$_7$. While the top (blue) and bottom (orange) spins make up two bulk all-in-all-out (AIAO) antiferromagnetic domains, individual tetrahedra with a one(blue)-in/three(orange)-out spin configuration form a domain wall with ferromagnetic projections perpendicular to the <1,1,1> plane. The local magnetization on each tetrahedron, either a three-in/one-out type or a two-in/two-out type, points along a local (1,1,1) or (1,0,0) axis, respectively, while magnetic domain walls exist in many orientations with minimal anisotropy [9-10, 22]. Hence the ferromagnetic moment can have both parallel and perpendicular projections to the domain wall. (B-C) Schematics of a two-rotator setup, including both a motorized continuous horizontal rotator, and a 3D-printed 15-degree-increment indexing stage. A plate-shaped sample was cooled through $T_N$ with a field-in-plane configuration (B), before it is flipped 90º to have the measurement field perpendicular to the surface at low temperature (C). The index stage sets the direction of the in-plane field cooling. (D) Representative odd-parity linear galvanomagnetic curves from two reciprocal Hall configurations and two vdP MR channels. Insets mark the lead configuration of each measurement. Between two reciprocal Hall configurations, $R_{43,12}(H=0) = -R_{21,43}(H=0)$, reflecting opposite geometric projections of MR channels. (E-G) Normalized linear slopes $1/\rho(H=0)$ d$\rho$/d$H$ as a function of in-plane **M** rotation angle $\phi$ for two Hall configurations, both above ($T$=300 K) and below ($T$=195 K) $T_N$= 227 K, and two vdP MR channels at 195 K. The $\phi$-averaged slopes for each of four channels are marked by dashed lines of the same color. The average values of two slopes of reciprocal Hall channels (purple diamonds) at $T$=195 K are also plotted against $\phi$, showing no angular dependence.

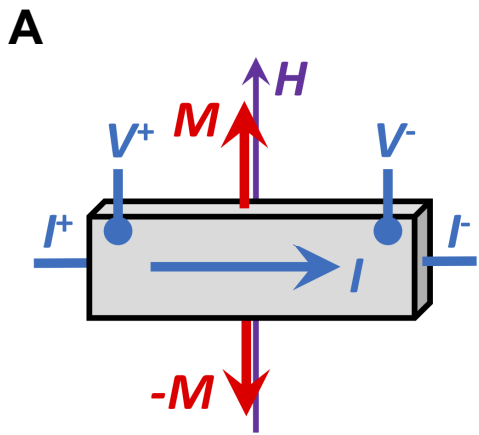
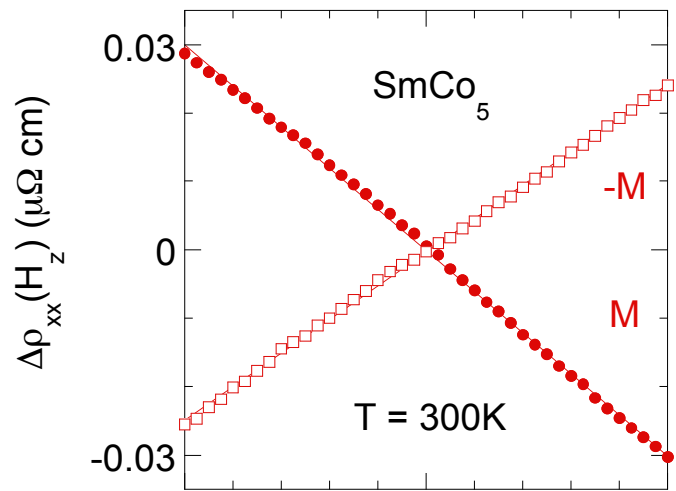
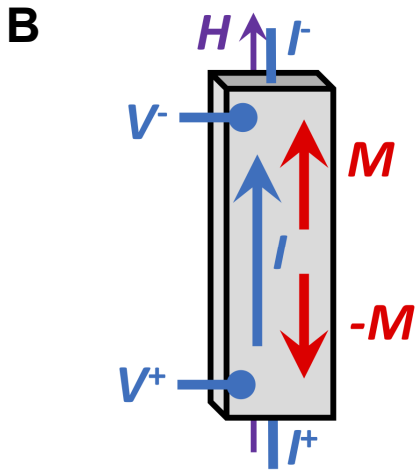
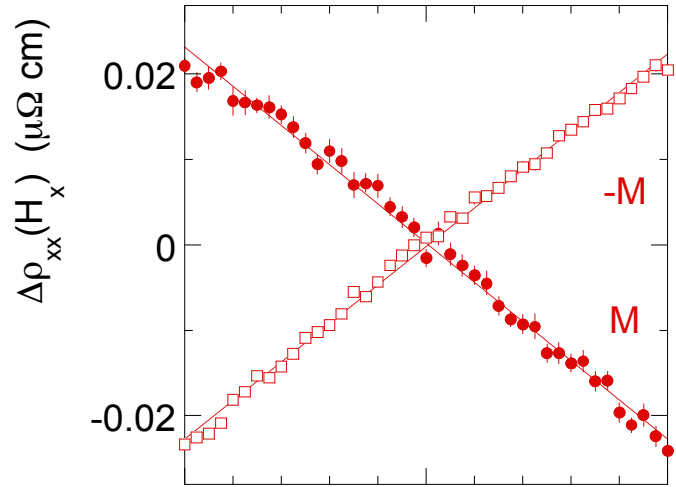
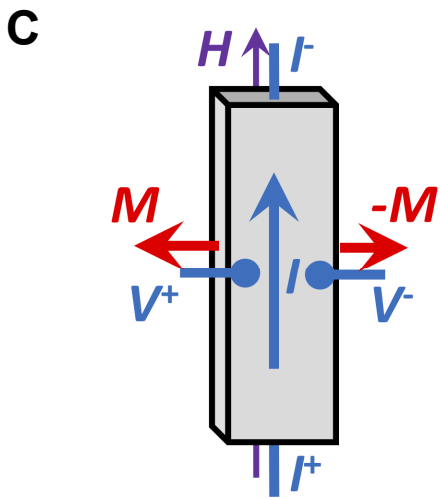
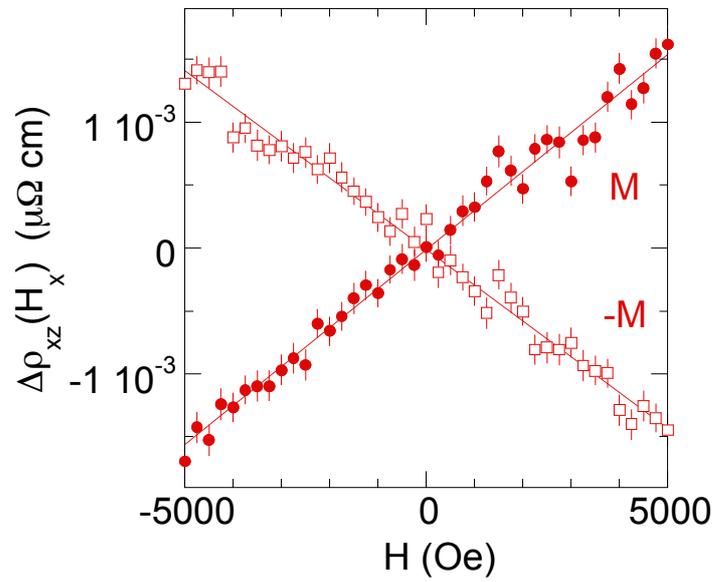

A

B

C

D

E

F

SmCo$_5$
T=300K

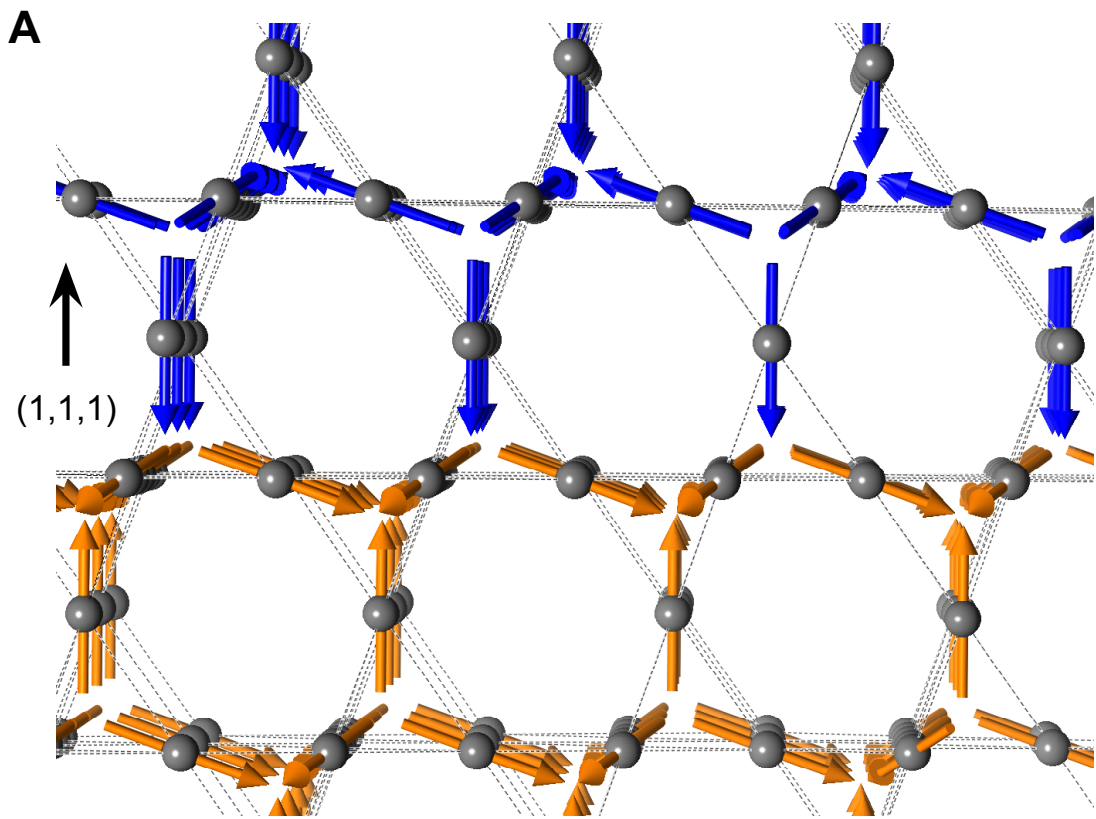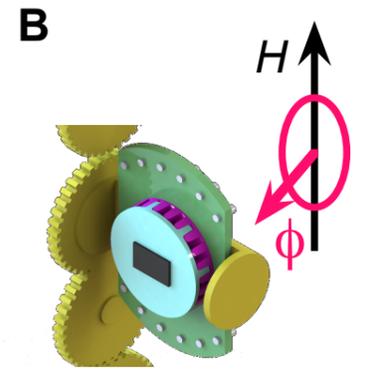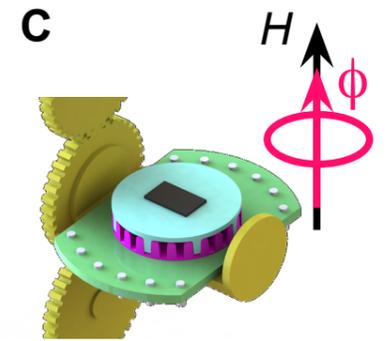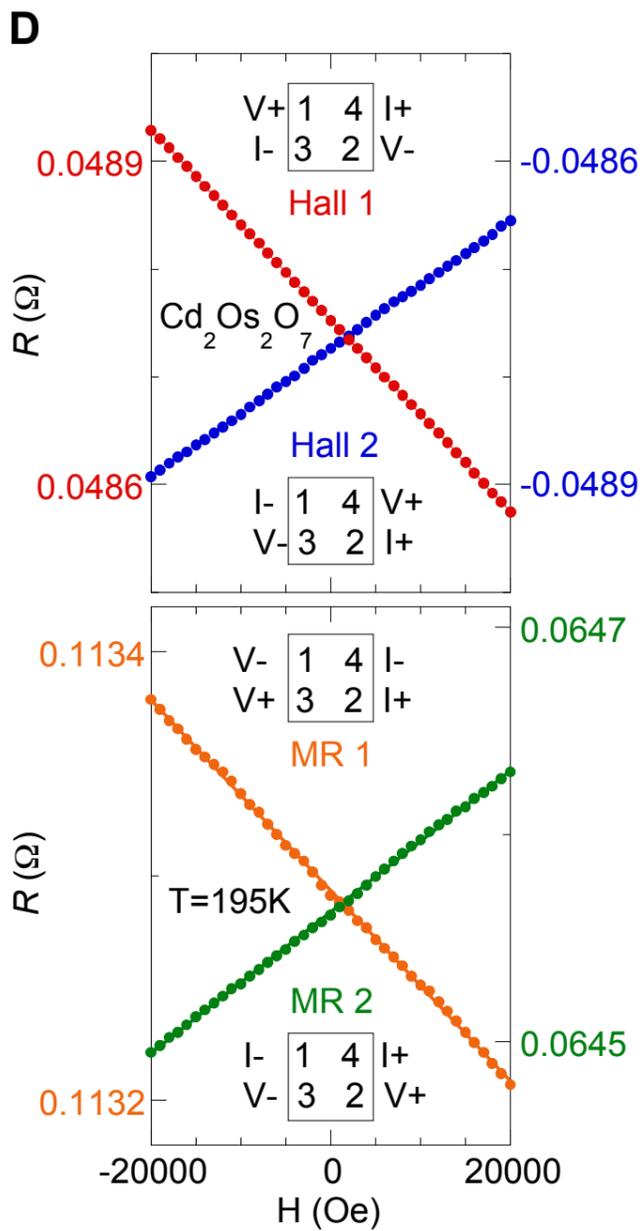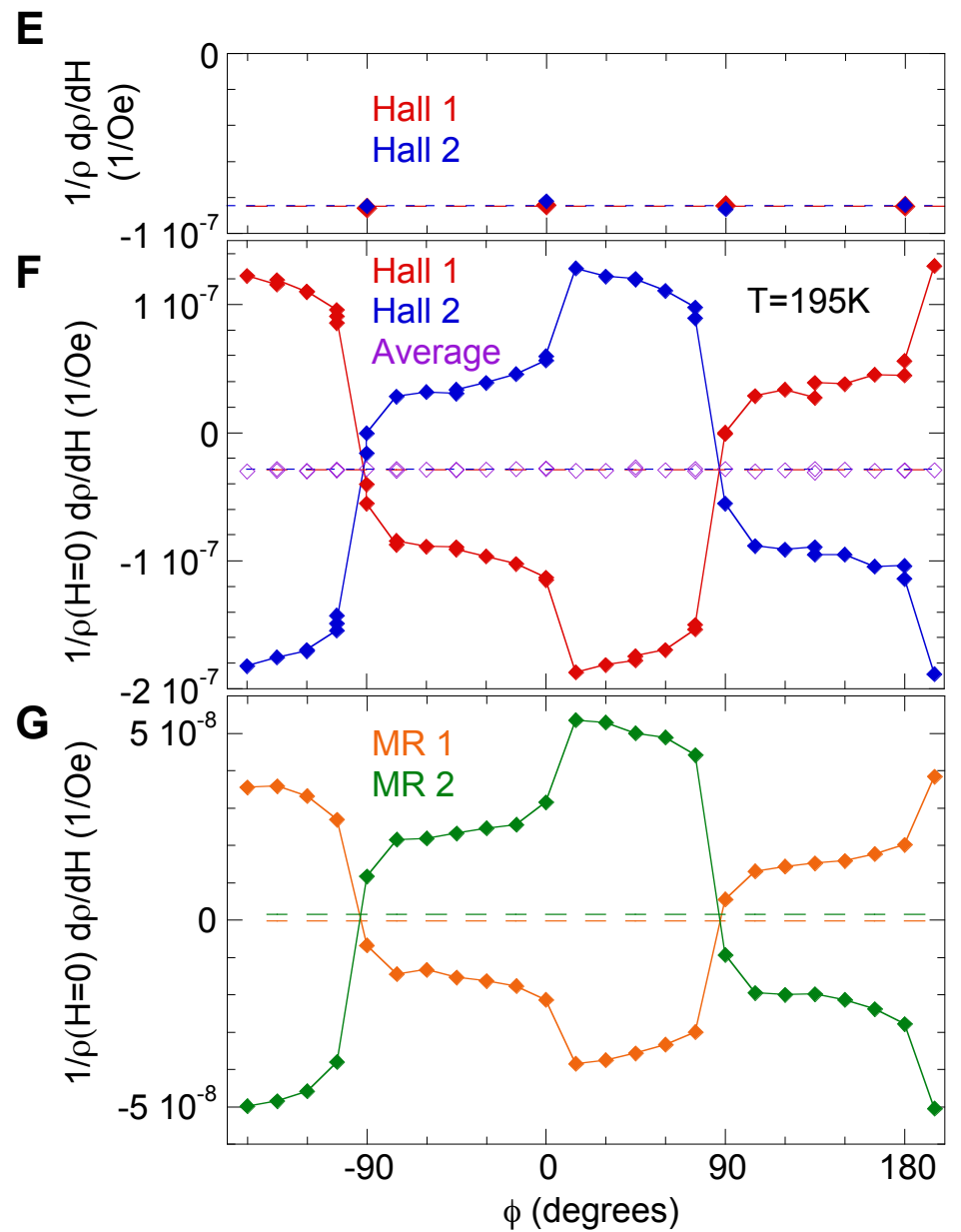

**Supplementary materials for:**

**Odd-parity linear magnetoresistance and the planar Hall effect**

Yishu Wang[1,2], Patrick A. Lee[1,3], D. M. Silevitch[1], F. Gomez[1], S. E. Cooper[4], Y. Ren[5], J.-Q. Yan[6], D. Mandrus[6,7], T. F. Rosenbaum[1,*], Yejun Feng[1,4,*]

**Materials and Methods:**

Commercial-grade, polycrystalline, samarium-cobalt and neodymium-iron-boron magnets were purchased from the McMaster-Carr Supply Company, USA. Electrical transport samples were sliced from the bulk and polished to bar shapes (typical size of 5×2×0.5 mm$^3$) with no further thermal processing nor chemical modification. X-ray measurements of the samarium-cobalt materials (see below) revealed a mixture of two phases, SmCo$_5$ with lattice constants of $a=b$=5.0122 Å, $c$=3.9774 Å, and Sm$_5$Co$_{19}$ with $a=b$=5.0387 Å, $c$=48.4955 Å. Molar percentages of the two phases are ~90% and ~10%, respectively. Both constituents are of hexagonal structure with the $c$-axis aligned along the magnetization direction, spanning a ±10-degree (FWHM) mosaicity (Fig. S1).

Cd$_2$Os$_2$O$_7$ single crystal samples of typical 3×3×1mm$^3$ size were grown by the vapor transport method, and possess a more significant number of grain boundaries by comparison to small octahedral-shaped crystals of 0.3-0.5 mm size [20]. After saw-dicing and polishing, plate-shaped samples of ~400 μm lateral size and 50 μm thickness, with a (1,1,0) surface normal, were wired up in a vdP geometry for galvanomagnetic measurements. The sample circuit was mounted with two rotational degrees of freedom, one provided by the horizontal rotator option of the Quantum Design PPMS, the second provided by a home-built, 3D-printed miniature indexing stage with 24 angular positions at 15-degree steps (Figs. 3B, 3C).

Crystalline structures of both samples were examined by hard x-ray (105.7 keV) diffraction at sector 11-ID-C of the Advanced Photon Source. Both samples were mounted in the x-ray transmission geometry to maximize sensitivity to their bulk properties. The diffraction data was collected by a two-dimensional Perkin Elmer amorphous silicon x-ray image plate.

Magnetization measurements of SmCo$_5$ and Cd$_2$Os$_2$O$_7$ were performed in either a 7T Magnetic Property Measurement System (MPMS3, Quantum Design) or a 9T Physical Property

Measurement System (PPMS DynaCool, Quantum Design) equipped with the Vibrating Sample Magnetometer option. Electrical transport data was obtained with a Lakeshore LS372 resistance bridge and a 3708 preamplifier, on samples mounted on a motorized horizontal rotator stage inside a 14T PPMS DynaCool. $SmCo_5$ samples typically were magnetized by a 14T field before measurement. Electrical transport measurements on $Nd_2Fe_{14}B$, similarly illustrating odd-parity linear MR in a highly coercive ferromagnet (Fig. S3), were performed using a Linear Research LR700 resistance bridge in a 9T PPMS DynaCool, using differently wedged blocks to vary angles between **M** and **H**.

**Supplementary Figure Captions:**

**Fig. S1. Magnetic and structural characteristics of SmCo.** (A) Magnetic hysteresis of the SmCo sample along the *c*-axis (easy axis), displaying a large hysteresis which increases as *T* is reduced from 300 K to 200 K. (B) Hard x-ray (105.7 keV) diffraction image of one of the transport samples. The diffraction data indicate an oriented polycrystal with the *c*-axis aligned to within ±10° FWHM mosaic range. Diffraction data also indicate that the alignment is totally random in the *a-b* plane of the tetragonal structure. (C) Radially integrated intensity of the diffraction pattern reveals a coexistence of two phases, $SmCo_5$ and $Sm_5Co_{19}$. Due to the textured structure, only unit cell parameters are refined.

**Fig. S2. Magnetic and structural characteristics of the $Cd_2Os_2O_7$ single crystal.** (A) For samples that are field cooled (4 T) through $T_N$, the magnetization (red) as a function of field is not hysteretic, but has a finite intercept at zero field. The frozen moment **M** changes sign when the field direction is reversed to -4 T in a separate cooldown (blue). (B) For a piece of $Cd_2Os_2O_7$ single crystal with a surface normal of (1,1,0), the zero-field frozen magnetization **M**(ϕ) was measured under various field cooldowns with a field-in-plane geometry of different ϕ. A finite **M** is always measured along the field direction, while the magnitude varies. (C) Hard x-ray (105.7 keV) diffraction image of the transport sample, showing the (1,1,0) zone with a two-fold symmetry. The spread of diffraction spots indicates a finite mosaic structure. Continuous circles are from x-ray diffraction of the rotator sample holder.

**Fig. S3. Odd-parity linear MR in $Nd_2Fe_{14}B$.** The highly coercive ferromagnet $Nd_2Fe_{14}B$ demonstrates odd-parity linear MR behavior similar to that of $SmCo_5$. The rotation test between **M**, **E**, and **H** is consistent with the scenario demonstrated in Fig. 2a, revealing both the (**M**·**H**)**E** and (**M**×**E**)×**H** terms.

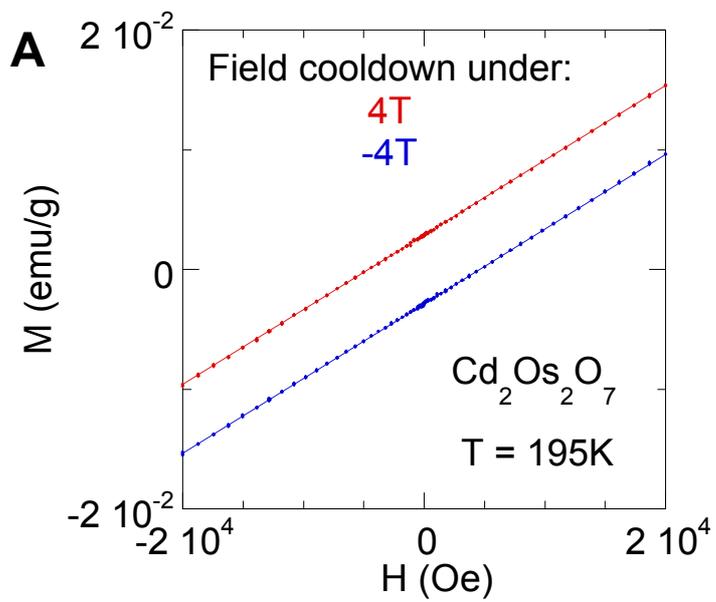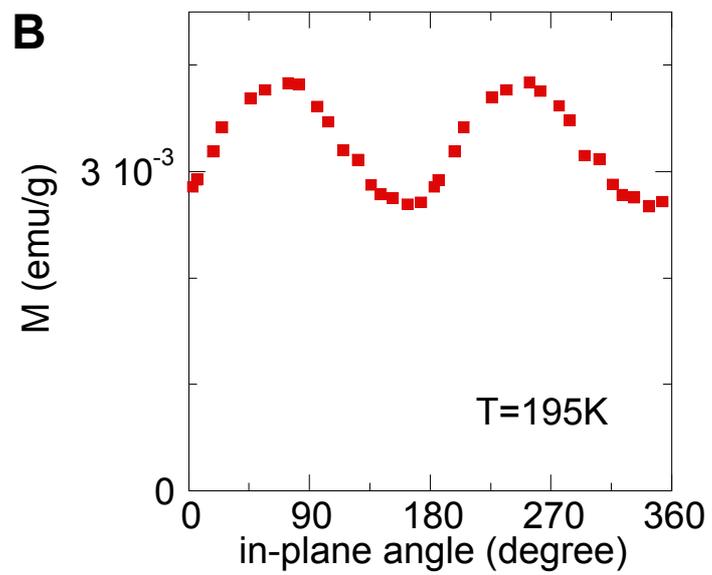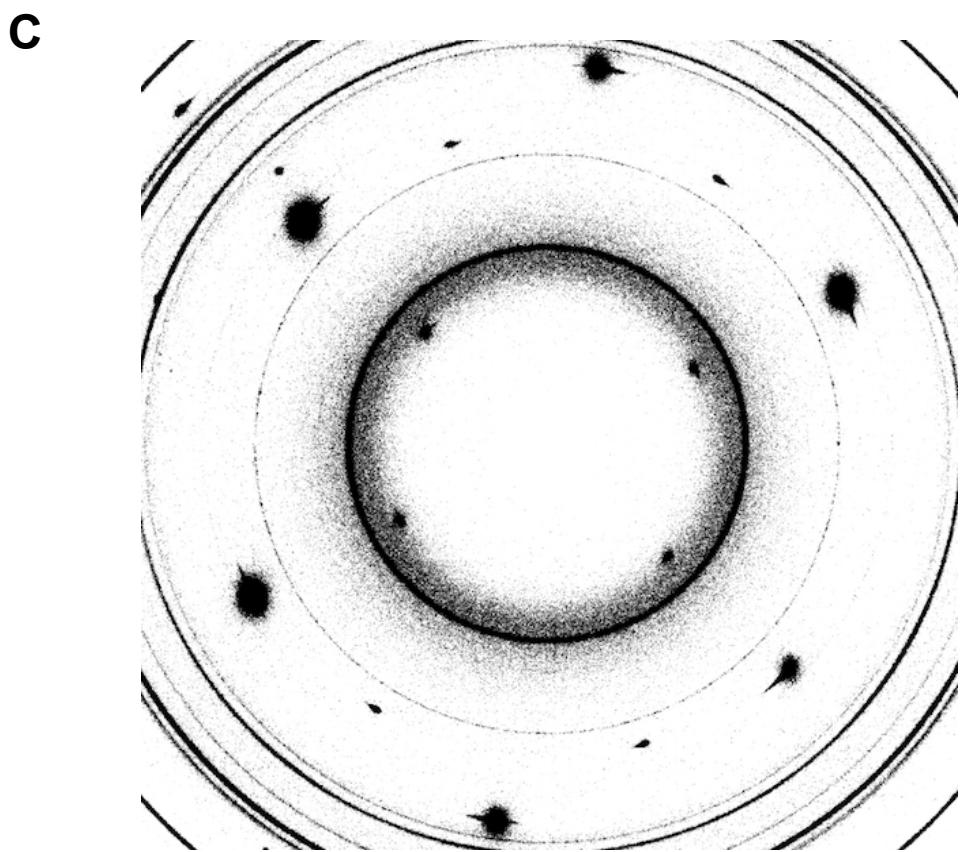

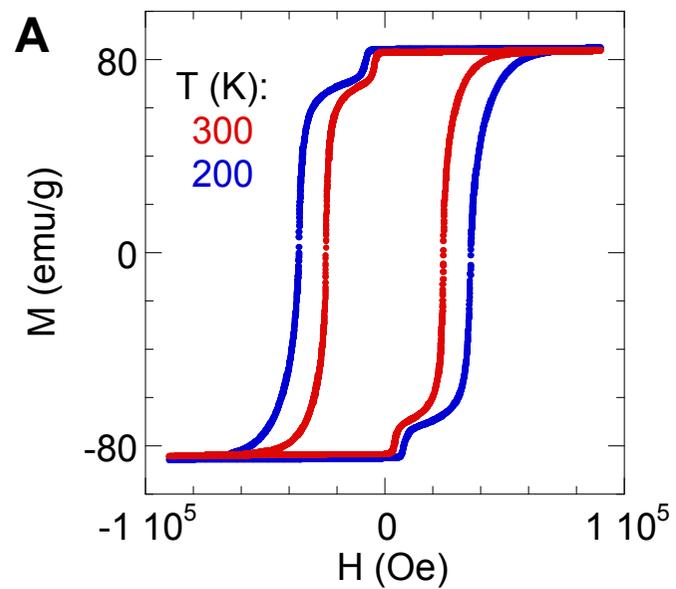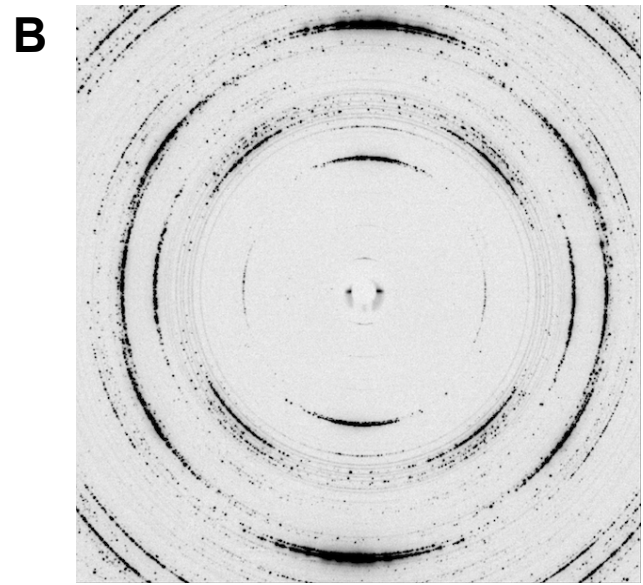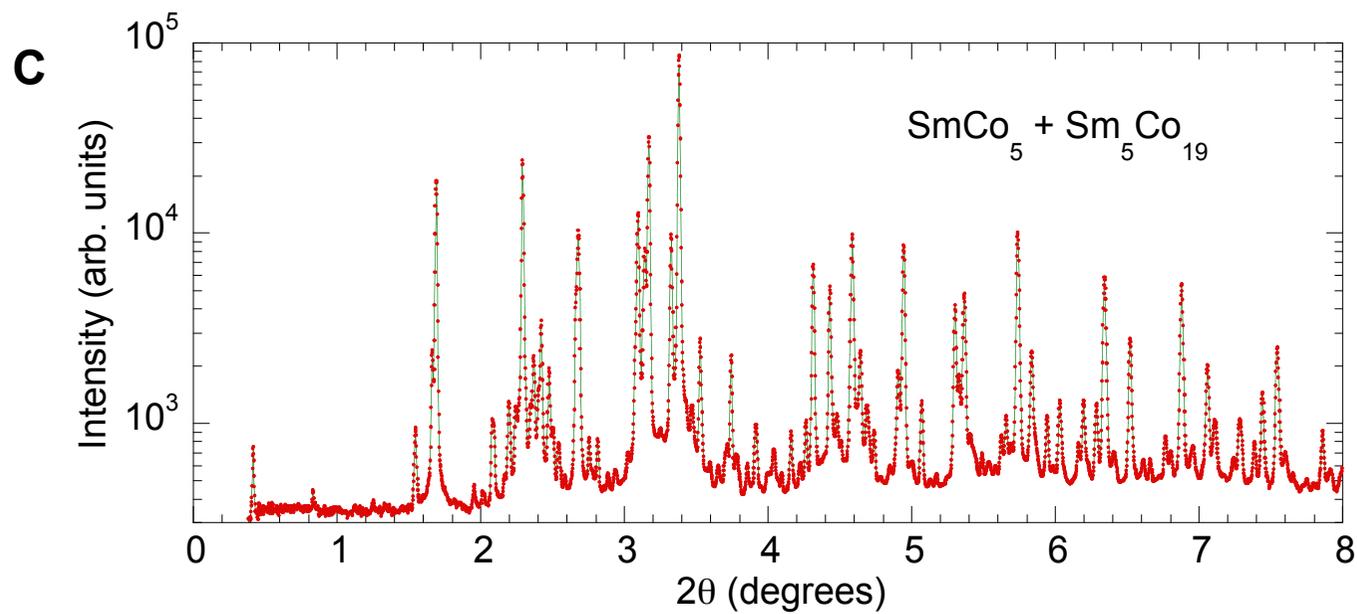

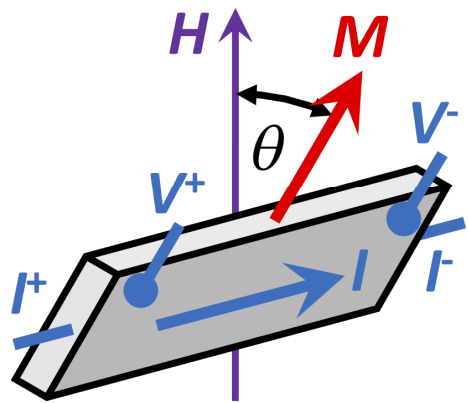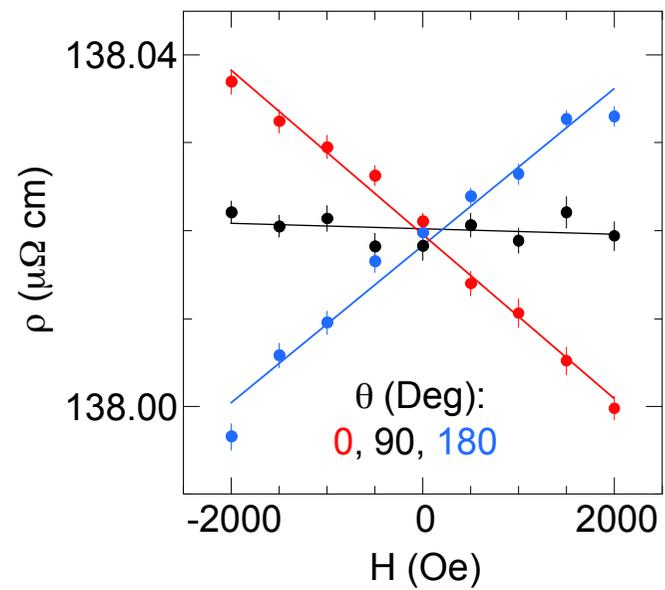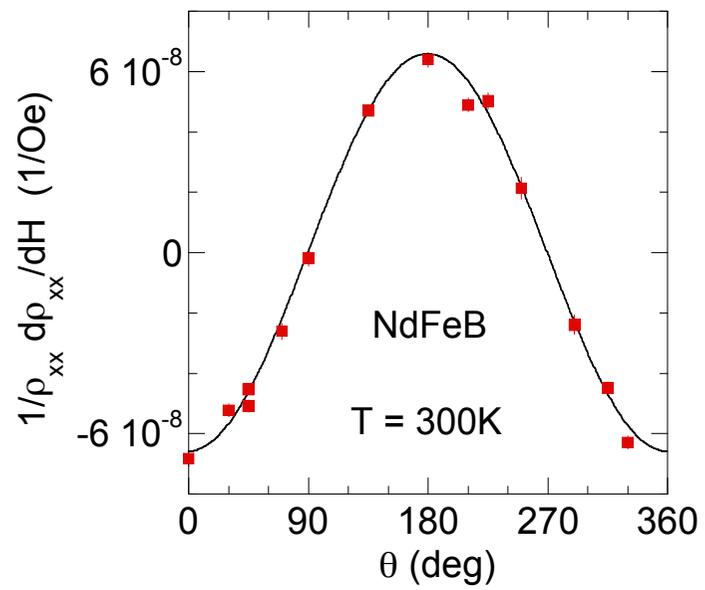